\documentclass[jkps,preprint,fleqn,nofootinbib,superscriptaddress,showpacs,showkeys]{revtex4}
\usepackage{amssymb}
\usepackage{amsmath}
\usepackage{bm}
\usepackage{epsfig}
\usepackage{graphicx}
\def\be{\begin{equation}} 
\def\ee{\end{equation}} 
\def\bea{\begin{eqnarray}}
\def\eea{\end{eqnarray}} 
\def\nnb{\nonumber}

\begin{document}
\title{
New Photodisintegration Model of GEANT4 for 
the $d\gamma \to np$ Reaction with a Dibaryon Effective Field Theory
}

\author{Jae Won Shin}
\affiliation{Department of Physics,
Soongsil University, Seoul 156-743, Korea}
\author{Chang Ho Hyun}
\email{hch@daegu.ac.kr}
\affiliation{Department of Physics Education, 
Daegu University, Gyeongsan 712-714, Korea}

\date{4 Aug. 2016}

\begin{abstract}

We develop a new hadronic model for GEANT4 
that is specialized for the disintegration
of the deuteron by photons, $d\gamma \to n p$.
For the description of two-nucleon interactions,
we employ a pionless effective field theory 
with dibaryon fields (dEFT).
We apply the new model of GEANT4 (G4dEFT) to the calculations of the total 
and the differential cross sections in $d\gamma \to n p$ 
and compare the results with empirical data.
As an application of the new model, 
we calculate the neutron yield from 
the $\gamma + {\rm CD}_2$ process.
G4dEFT predicts peaks for the neutron yield, 
but the existing model of GEANT4 does not show such behavior.

\end{abstract}


\pacs{24.10.Lx, 25.20.-x, 13.60.-r}

\keywords{$d\gamma \to np$, Effective field theory, Dibaryon fields, GEANT4}

\maketitle

\setcounter{page}{1}

\section{Introduction}

Effective field theory (EFT) 
has become a popular method to study hadronic reactions 
with and without external probes at low energies.
In the low-energy region where
the scale of momenta is much smaller than the mass of pions, 
treating the exchange of pions, 
as well as heavy degrees, 
in terms of effective contact interactions 
may be reasonable. 
Theories thus constructed are called 
pionless effective field theory \cite{pionless_1,pionless_2,pionless_3,pionless_4,pionless_5,pionless_6,pionless_7}. 
In this work, 
we employ a modified version of pionless theory, 
so-called pionless theory, with dibaryon fields (dEFT) \cite{deft_1}.
In the dEFT, 
a perturbative expansion is directly manifested in 
physical observables such as 
cross-sections and their analytic forms are available. 

Particle transport codes such as GEANT4 (GEometry ANd Tracking) \cite{g4n1, g4n2} 
are commonly utilized in the setup of plans 
for efficient experiments and in the analysis of results. 
Doing so, we can accurately estimate observables 
for the reactions under consideration. 
However, one of the authors (JWS) recently reported that 
built-in hadronic models of GEANT4 (v10.0) 
fail to describe peaks that are produced 
through the $^{9}$Be(p,n)$^{9}$B reaction in the low-energy region \cite{g4dce}. 
GEANT4 was originally built for high-energy physics.
In some cases, 
it fails to describe low-energy phenomena correctly as was shown 
in the $^{9}$Be(p,n)$^{9}$B reaction.
In Ref. 11, 
the authors developed 
a new data-based charge-exchange model for
the $^9$Be(p,n)$^9$B reaction,
and were able to achieve good agreement with experimental data.

Inspired by the observation in Ref. 11, we extend the application of GEANT4
models to low-energy processes in the two-nucleon systems.
We simulate the $d\gamma \to np$ reaction by using GEANT4 code 
and find that an existing model of GEANT4 (v10.1) gives a null result for
the total cross-section of the $d\gamma \to np$ reaction 
at energies below the pion threshold.
In order to obtain a better description 
at energies below the pion threshold, 
we construct a dEFT-based hadronic model of GEANT4 for the $d\gamma \to np$ reaction. 
To check the validity of our model, 
we simulate the total and the differential cross-sections of the $d\gamma \to np$ process
by using GEANT4 with dEFT (G4dEFT) at various energies and angles,
and we compare the results with available experimental data.
The result proves the usefulness of the combination of dEFT with GEANT4.
As an application of the model, we calculate the neutron yields from the
$\gamma +\rm{C} \rm{D}_2$ reaction.
We find that the G4dEFT model predicts results critically 
different from those predicted by using 
existing GEANT4 models.

We organize the paper as follows;
In Section~II, we briefly 
present analytic formulae for the differential cross-section of the dEFT, 
and the description of GEANT4 models relevant to this work. 
In Section~III, we present the results for the cross sections in $d\gamma \to np$ 
with G4dEFT and compare them with experimental data.
We propose experiments from which we can test the 
predictive power of G4dEFT.
We summarize the work in Section~IV.

\section{Methods
\label{sec:meth}}


Large scattering lengths signal the presence of 
weakly-bound or almost-bound states. 
Through the inclusion of dibaryon fields that represent 
a weakly-bound state ($^3 S_1$ channel) or an almost-bound state ($^1 S_0$ channel) of two nucleons, 
this formulation simplifies calculations compared to the pionless theories without dibaryon fields \cite{bedaque, kolck, deft1}. 
We showed that the dEFT up to next-to-leading order (NLO) could be 
applied to diverse two-nucleon systems successfully at low energies \cite{deft1, deft2, deft3, deft4, deft5, deft5b, deft6}. 

The differential cross-section for the $d \gamma \to np$ reaction 
can be written as 
\begin{eqnarray}
\frac{d \sigma}{d \Omega} = \frac{\alpha}{24 \pi}\frac{pE_{1}}{k} \sum_{\rm spin} |A|^{2}, 
\label{eq:difcs}
\end{eqnarray}
where $\alpha$ is the fine structure constant, 
$k$ is the energy of an incoming photon, and 
$E_{1} = \sqrt{m^{2}_{N} + p^{2}}$ is the energy 
of an outgoing nucleon in the center-of-mass (c.m.) frame. 
The spin-averaged square of the amplitude 
can be written as
\begin{eqnarray}
S^{-1} \sum_{\rm spin} |A|^{2} &=& 16(|X_{MS}|^{2}+ |Y_{MV}|^{2}) 
+ 8(|X_{MV}|^{2} + |Y_{MS}|^{2}) 
\nnb \\ 
&+& 12[1-(\hat{k}\cdot \hat{p})^{2}](|X_{E}|^2+|Y_{E}|^{2}),
\end{eqnarray}
where $S$ is a symmetry factor for the spin average, $S=2$. 
The matrix elements $X_{MV}$, $X_{MS}$, $X_E$, $Y_{MV}$, $Y_{MS}$, $Y_E$, and 
the definitions of $\hat{k}$ and $\hat{p}$ 
can be found in Ref. 20. 


GEANT4 (GEometry ANd Tracking) 
is a simulation tool kit written in the C++ language, 
which allows microscopic Monte Carlo simulations of the
propagation of particles interacting with materials.
It is being widely and successfully used in many different scientific fields, 
such as neutron shielding studies \cite{g4nshield1, g4nshield2}, 
medical physics \cite{G4Med2, Shin1, Shin1a}, 
accelerator-based radiation studies \cite{ShinAcc1, Shin2, g4dce, g4Kirams},
environment radiation detection studies \cite{G4Det2, G4Det3, Shin4}, etc.

There are many GEANT4 physics models and cross-sections 
for both hadronic and electromagnetic~(EM) interactions. 
For photonuclear interactions, however, 
the only GEANT4 cross-section and hadronic model 
relevant to this work are 
the ``G4PhotoNuclearCrossSection" class \cite{g4_emExtra} 
and the ``G4CascadeInterface" \cite{BERTIref3} class, respectively.
The G4PhotoNuclearCrossSection gives the total inelastic 
cross-sections for photonuclear interactions. 
A Bertini-style cascade model is used for G4CascasdeInterface 
to calculate the final states of photonuclear reactions. 
The Bertini model is available for the incident photon energies
0 $\leq$ $E_{\gamma}$ $<$ 3.5 GeV 
in GEANT4 (v10.1), 
but the model has been tested mostly 
in the energy range 60 MeV $\sim$ 3 GeV \cite{BERTIref3}. 
GEANT4 uses them by using the ``G4EmExtraPhysics" GEANT4 
Physics Constructor. 

\section{Results}

\subsection{Total Cross-Section
\label{sec:totalcros}}

\begin{figure}[tbp] 
\centering
\epsfig{file=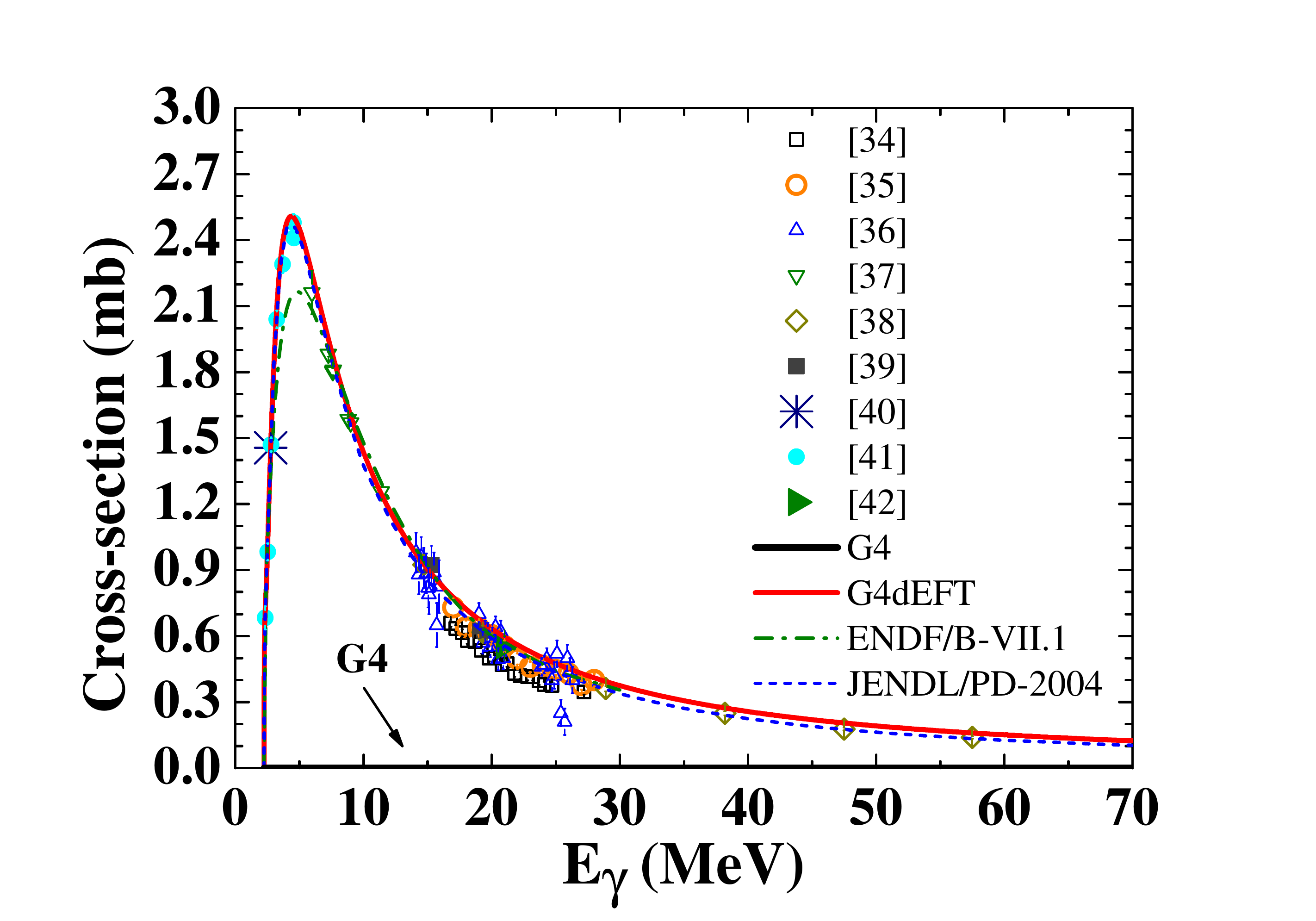, width=10cm}
\caption{(Color online) Total cross-section of the $d \gamma \to np$ reaction
with respect to the incident photon energy in the laboratory frame ($E_\gamma$).
Symbols denote the experimental data 
\cite{totE_1, totE_2, totE_3, totE_4, totE_5, totE_6, totE_7, totE_8, totE_9} 
taken from the EXFOR database \cite{exfor}. 
Dot-dashed and dashed lines represent the nuclear data extracted from ENDF/B-VII.1 \cite{endf} and 
JENDL/PD-2004 \cite{jendl}, respectively.
Black and red solid lines are the results obtained from G4 and G4dEFT, respectively. 
}
\label{figure2}
\end{figure}

To check the validity of GEANT4 for the $d\gamma \to np$ reaction, 
we first calculate the total cross-section of the reaction by using G4EmExtraPhysics (G4). 
In Fig.~\ref{figure2}, we show the result of G4 as a black solid line,
which is consistently zero in the energy range $E_\gamma \leq 70$ MeV.
Evidently, the simulation using GEANT for the few-nucleon
systems at low energies must be improved. 
We try to incorporate the formulae of dEFT.

Figure~\ref{figure2} shows the total cross-section 
calculated by using G4dEFT (red solid line), 
together with the experimental data \cite{totE_1, totE_2, totE_3, totE_4, totE_5, totE_6, totE_7, totE_8, totE_9}, 
and the evaluated nuclear data from ENDF/B-VII.1 \cite{endf} and JENDL/PD-2004 \cite{jendl}. 
ENDF/B-VII.1 does not reproduce some experimental data around the peak, 
but our G4dEFT result shows good agreement 
with data not only around the peak, but also over the energy range considered.

\begin{figure}
\begin{center}
\epsfig{file=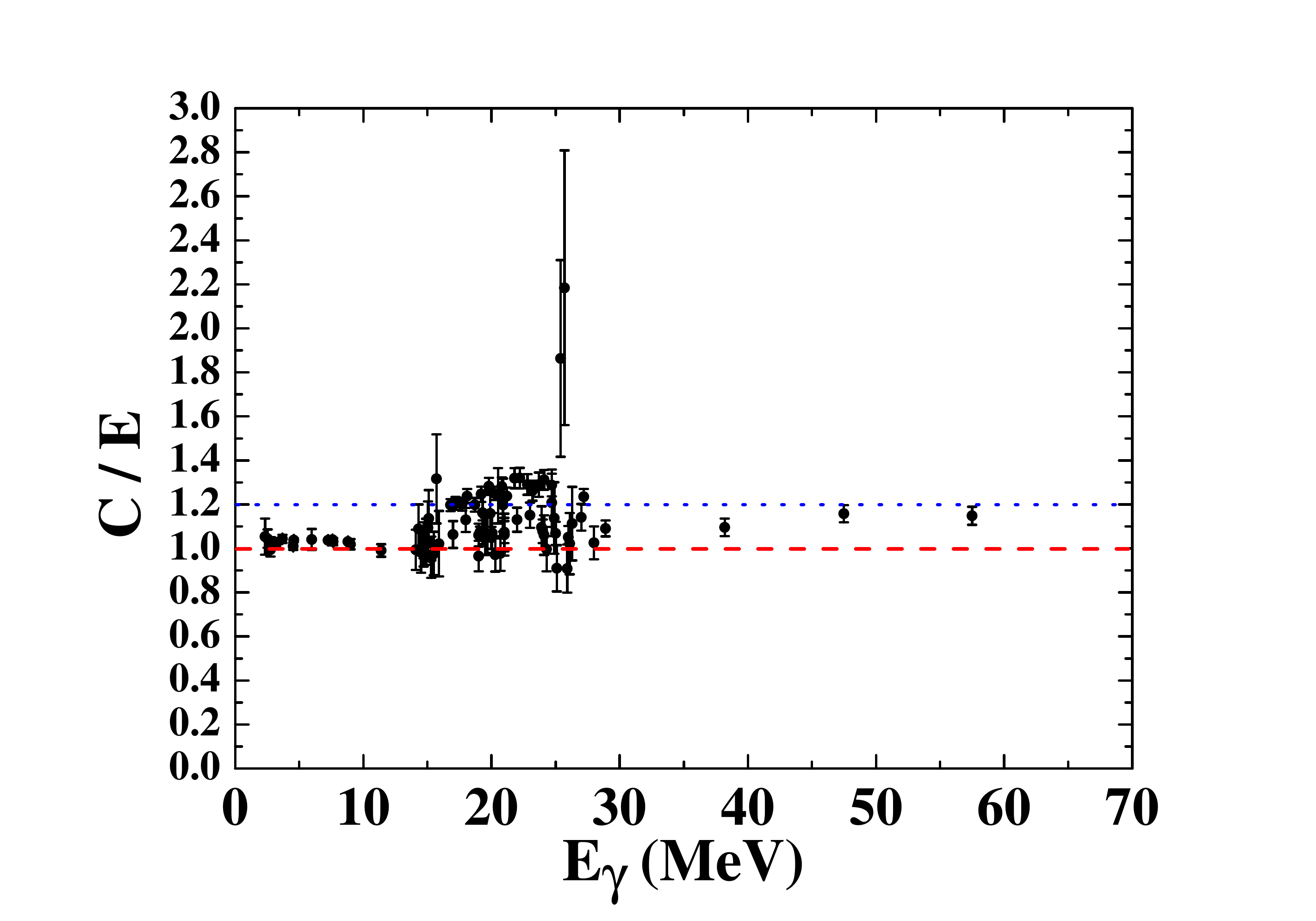, width=10cm}
\end{center}
\caption{
Ratio of the calculated total cross-section 
for the $d\gamma \to np$ reaction 
obtained by using G4dEFT to the experimental data 
\cite{totE_1, totE_2, totE_3, totE_4, totE_5, totE_6, totE_7, totE_8, totE_9}.
}
\label{fig:cofe}
\end{figure}

The ratios of the total cross sections obtained by using G4dEFT (denoted by C) 
to those from experiments (denoted by E)
\cite{totE_1, totE_2, totE_3, totE_4, totE_5, totE_6, totE_7, totE_8, totE_9} are shown
Figure~\ref{fig:cofe}. 
The results from G4dEFT are consistent with the experimental data 
for the photon energy $E_\gamma$ below 15 MeV within experimental errors.
For the region 15 $<$ $E_\gamma$ $<$ 30 MeV, 
the ratios are largely located in the range from 1.0 to 1.2. 
There are large discrepancies (factors of $\sim$ 2) for 
$E_\gamma = 25.4$ MeV and $E_\gamma = 25.7$ MeV, but experimental errors are also large.
For the photon energy $E_\gamma$ above 30 MeV, 
four experimental data are shown and agree with those from G4dEFT 
within about 20\% error. 
We note that, in principle, the pionless theory will break down 
for momenta larger than the mass of pion. 
In this respect, it may make sense to compare the result with data
for the photon energies below about 20~MeV. 
One can see that the calculated results agree
well with the data in this region.

\subsection{Differential Cross-Sections}

In order to investigate the quality of the model in more detail,
we compare the differential cross-sections obtained from G4dEFT with
the experimental data available in literature or the EXFOR database \cite{exfor}. 
\begin{figure}[tbp] 
\centering
\epsfig{file=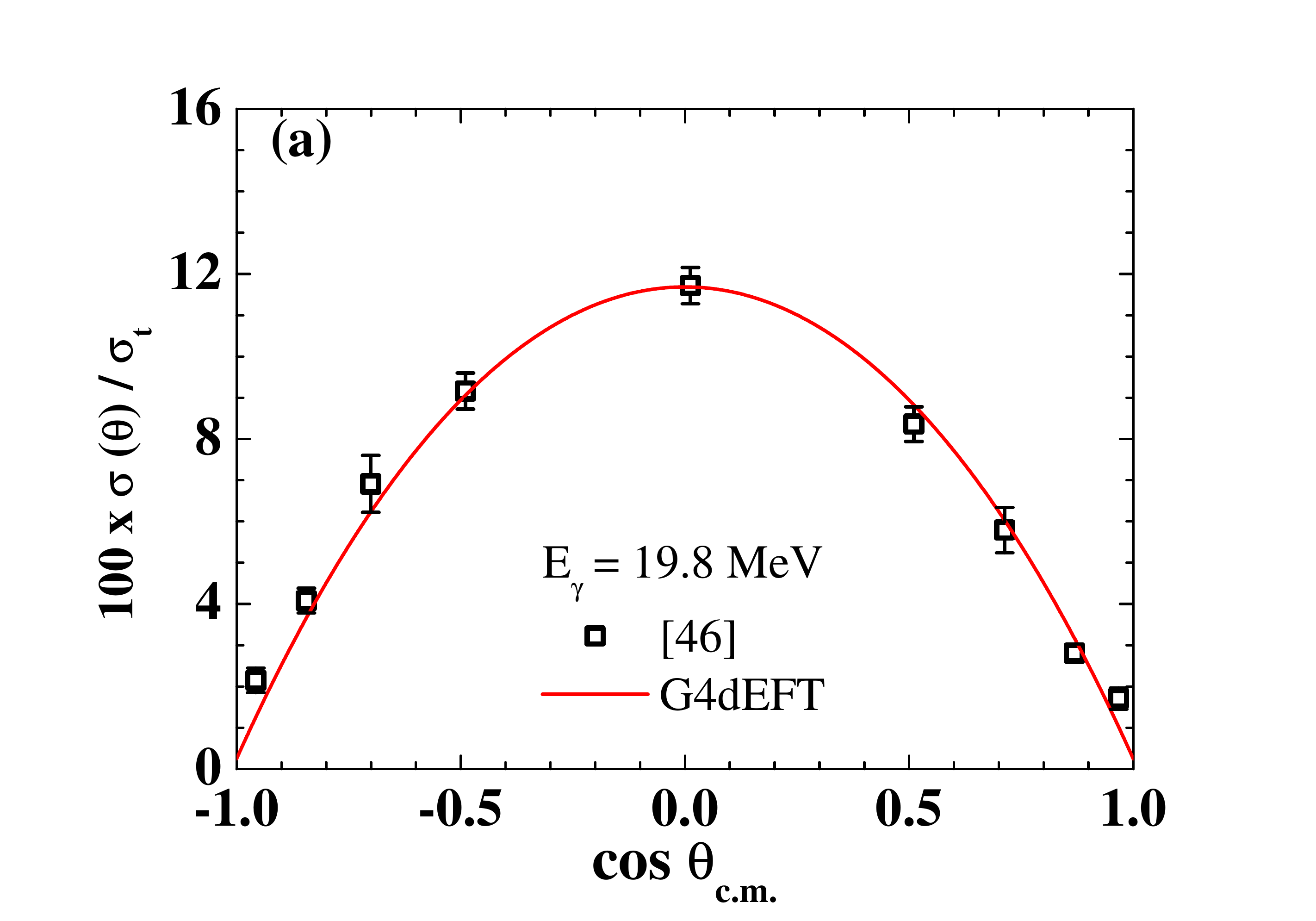, width=8cm}
\epsfig{file=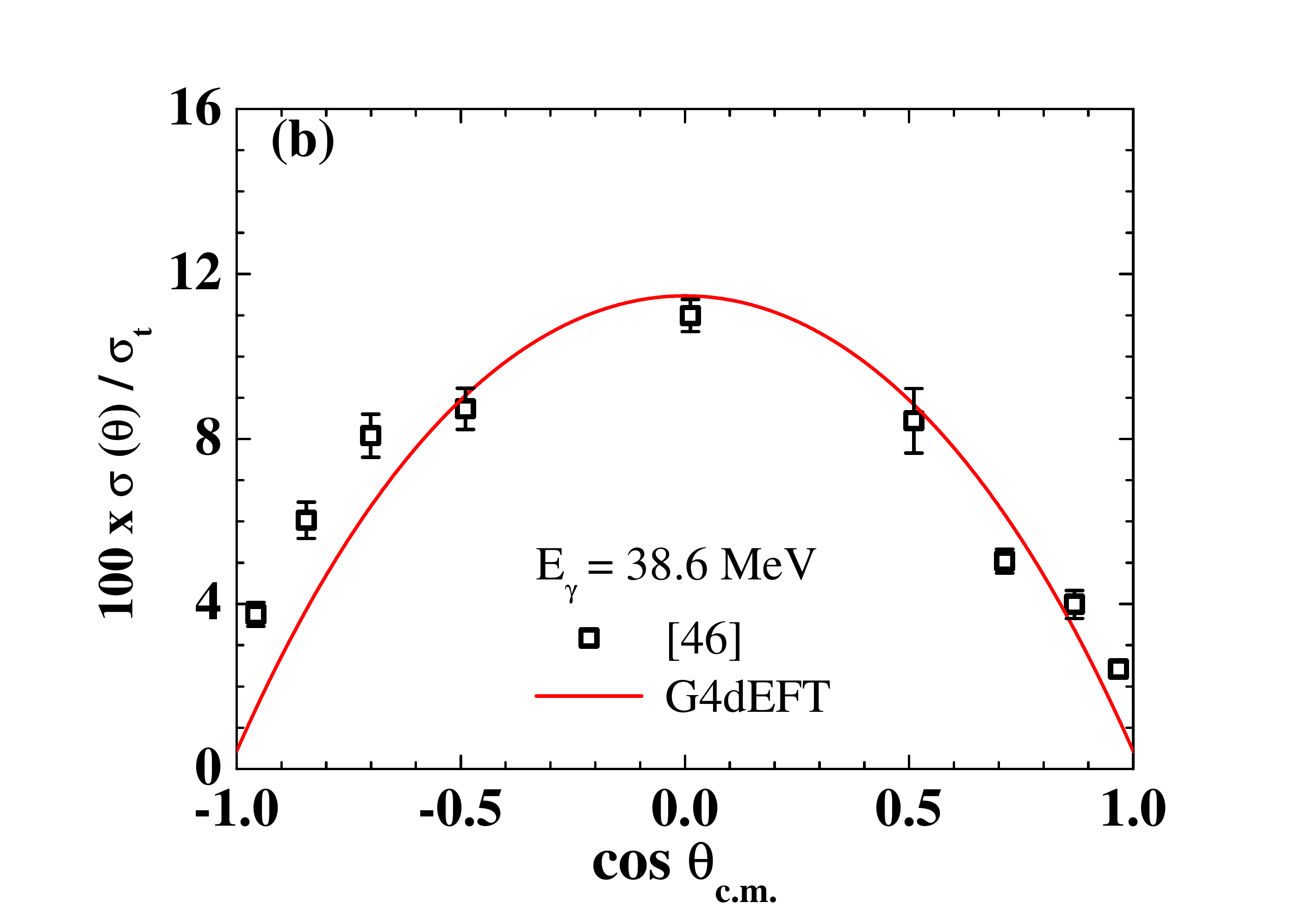, width=8cm}
\caption{(Color online) 
Differential cross-section $\sigma(\theta)$ as a function of $\cos \theta$ multiplied
by a factor $100/\sigma_t$, 
where $\sigma_t$ denotes the total cross-section of the unpolarized
photons at the energies
(a) $E_{\gamma} = 19.8$ MeV and (b) $E_{\gamma} = 38.6$ MeV. 
The open squares represent the experimental data 
taken from Ref. 46, and 
the solid lines denote the results obtained from G4dEFT.}
\label{figure4}
\end{figure}
In Fig. \ref{figure4}, 
we present the differential cross-sections at photon energies
(a) $E_\gamma = 19.8$ MeV and (b) $E_\gamma = 38.6$ MeV as functions 
of the polar angle $\theta$ in the c.m. frame.
Open squares are the experimental data from 
Ref. 46, 
and solid lines represent the results of G4dEFT. 
For $E_\gamma = 19.8$ MeV, G4dEFT agrees 
with the experiment data within the error bars, 
but for $E_\gamma = 38.6$ MeV, 
there are systematic underestimates in the backward angles.

\begin{figure}[tbp] 
\centering
\epsfig{file=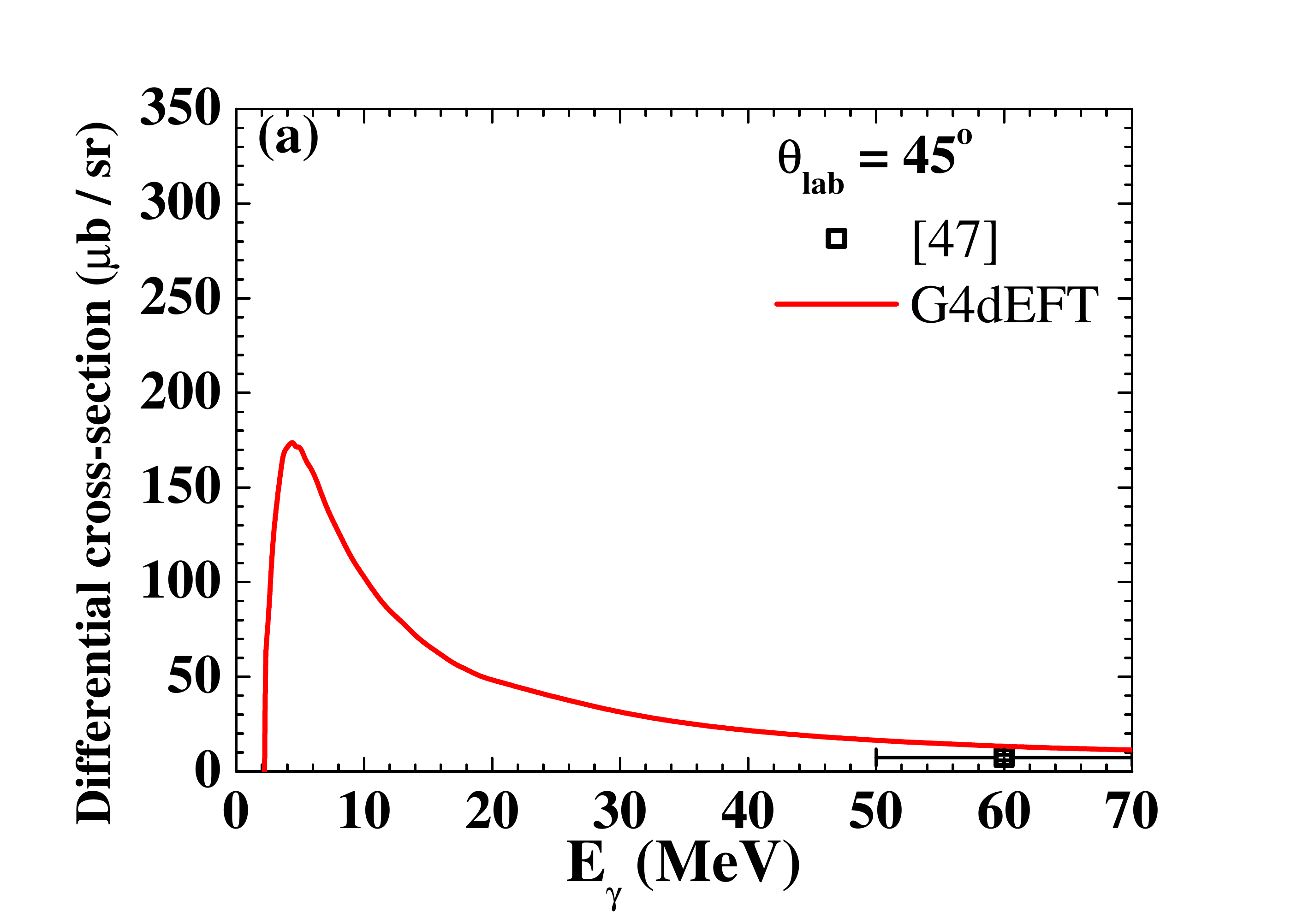, width=2.5in}
\epsfig{file=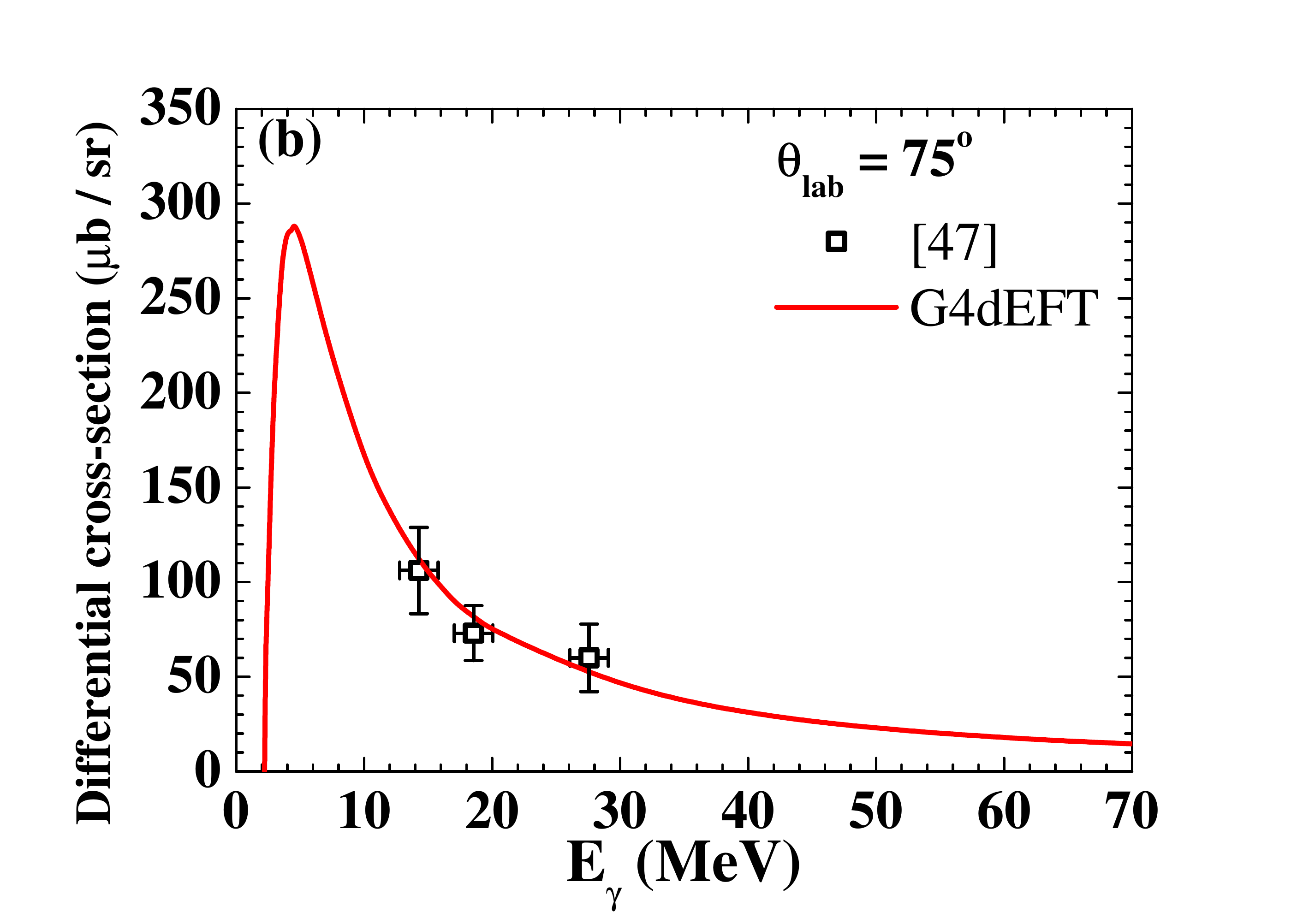, width=2.5in}
\epsfig{file=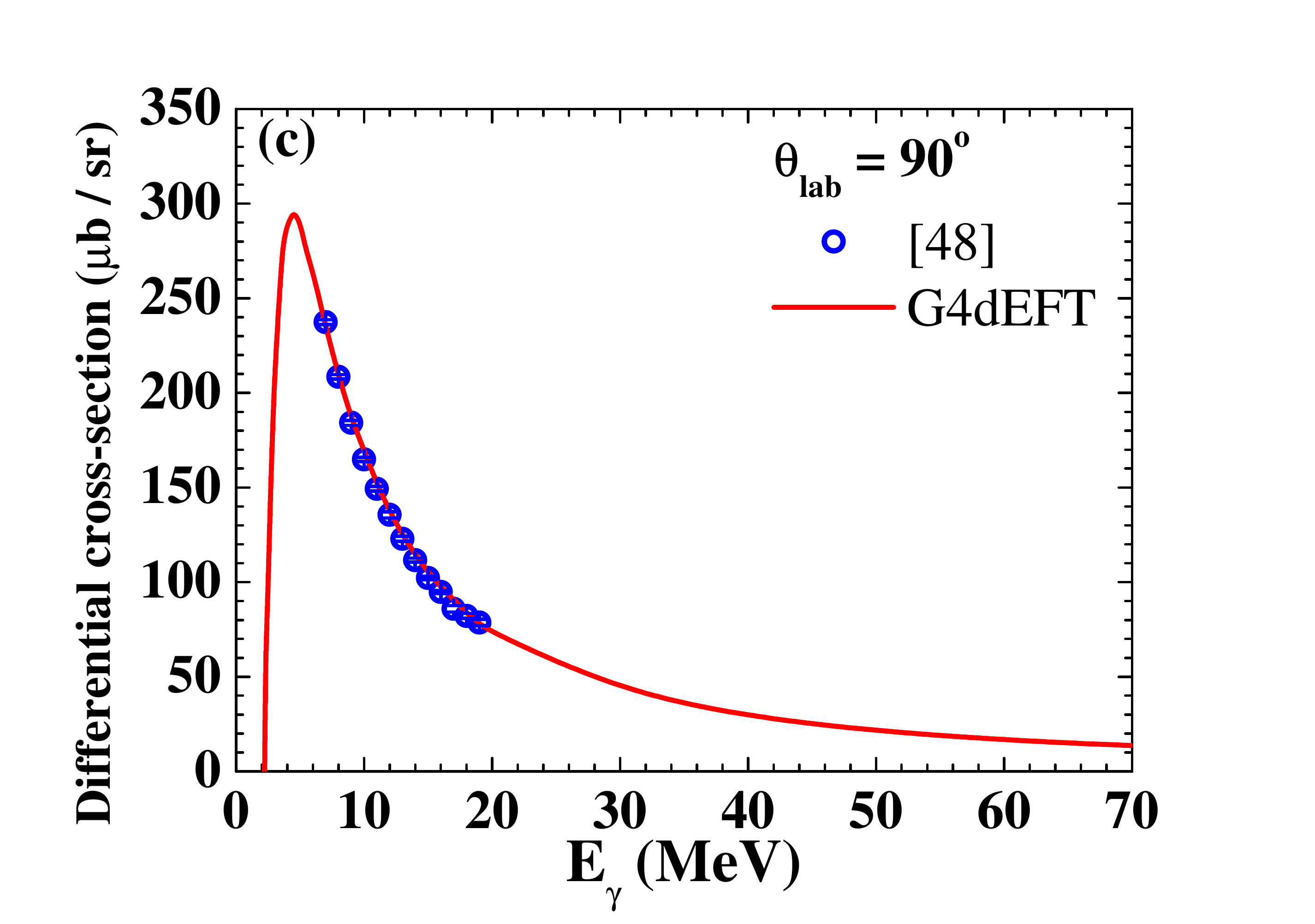, width=2.5in}
\epsfig{file=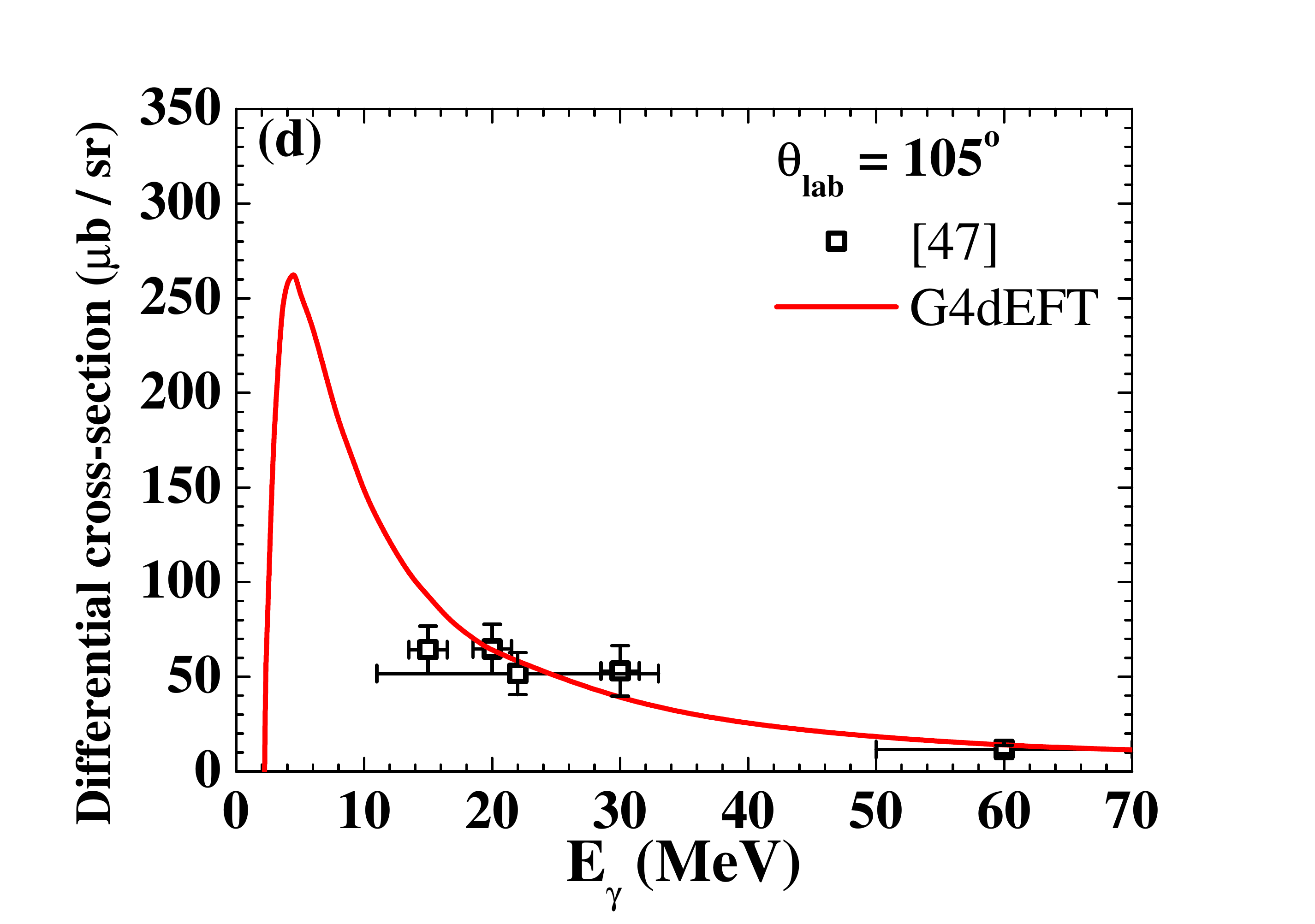, width=2.5in}
\caption{(Color online) 
Differential cross-sections 
at $\theta_{\rm lab} = 45^{\circ}, 75^{\circ}, 90^{\circ}$ 
and $105^{\circ}$. 
Open squares and circles correspond to the experimental data 
in Ref. 47 and Ref. 48, respectively,
and the values are taken from the EXFOR database \cite{exfor}.
Solid lines are the results obtained from G4dEFT.}
\label{figure5}
\end{figure}

Figure \ref{figure5} shows 
the differential cross-sections at 
$\theta_{\rm lab} = 45^\circ, 75^\circ, 90^\circ$ and $105^\circ$
as functions of the incident photon energy $E_\gamma$, 
where $\theta_{\rm lab}$ is the polar angle in the laboratory frame.
Open squares and circles represent the experimental data 
taken from Ref. 47 and Ref. 48, respectively. 
Solid lines denote the results obtained from G4dEFT. 
In the energy range $E_\gamma < 30$ MeV, 
the results of G4dEFT are mostly within the error bards 
regardless of the detection angles.

\begin{figure}[tbp] 
\centering
\epsfig{file=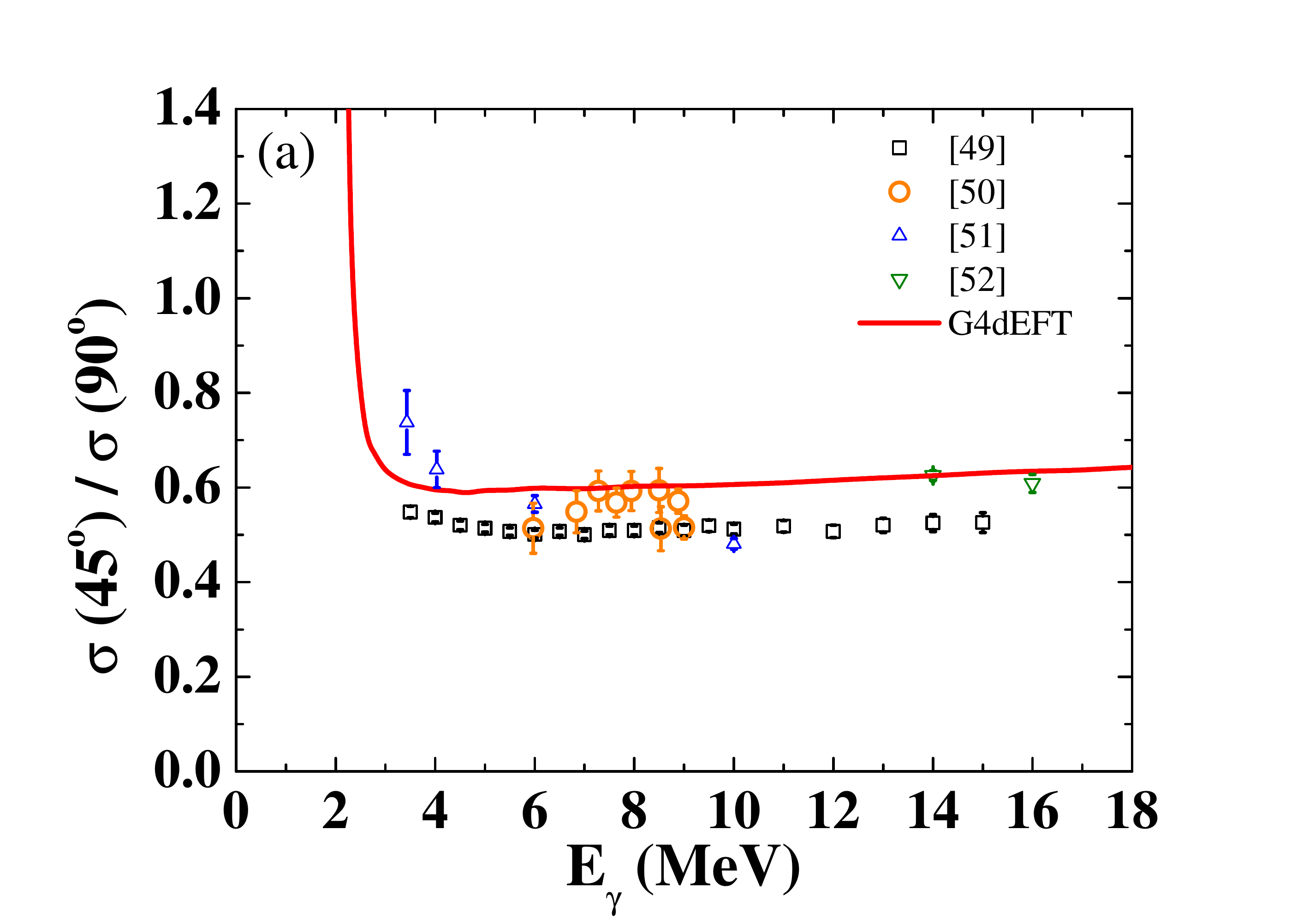, width=8cm}
\epsfig{file=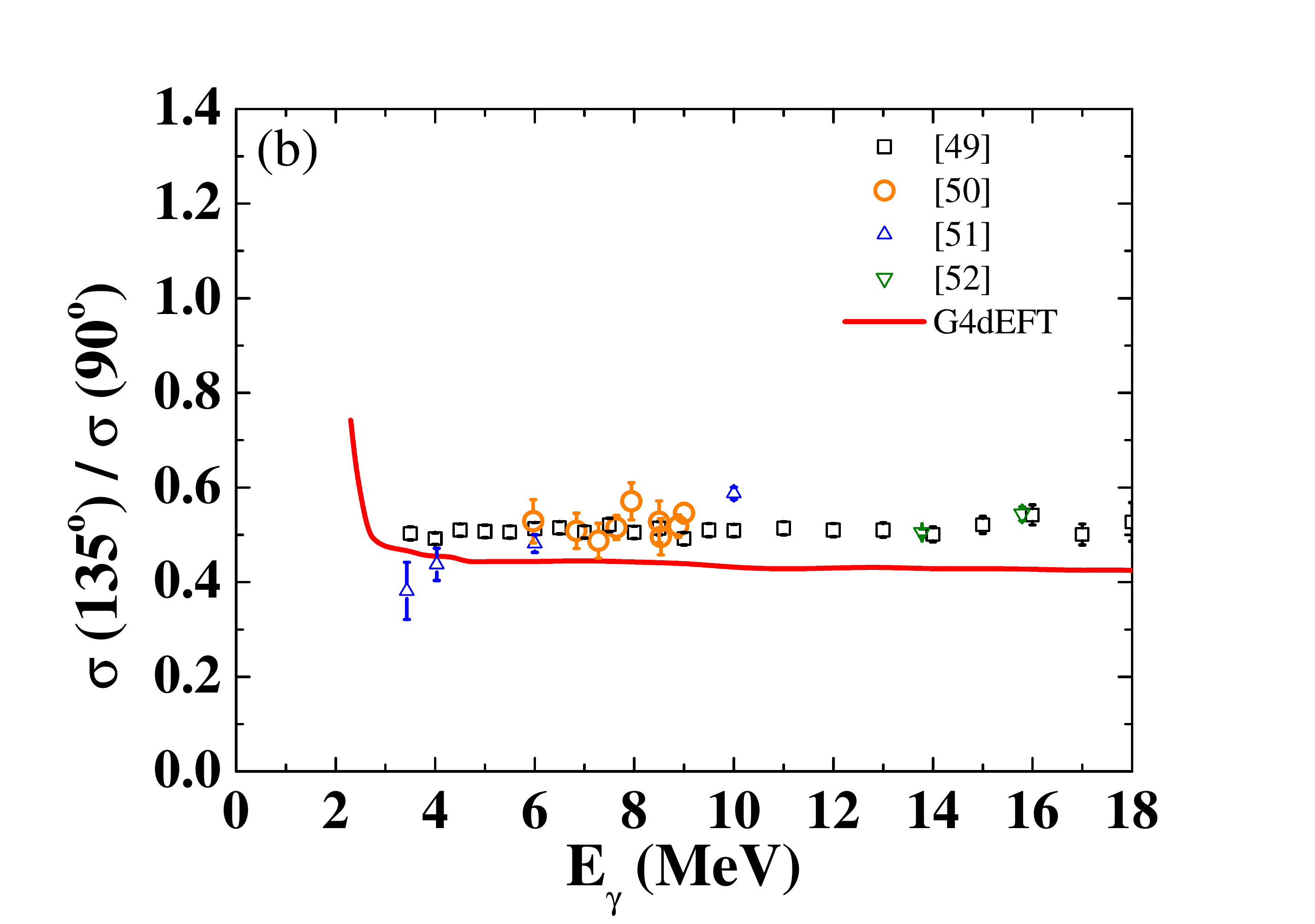, width=8cm}
\epsfig{file=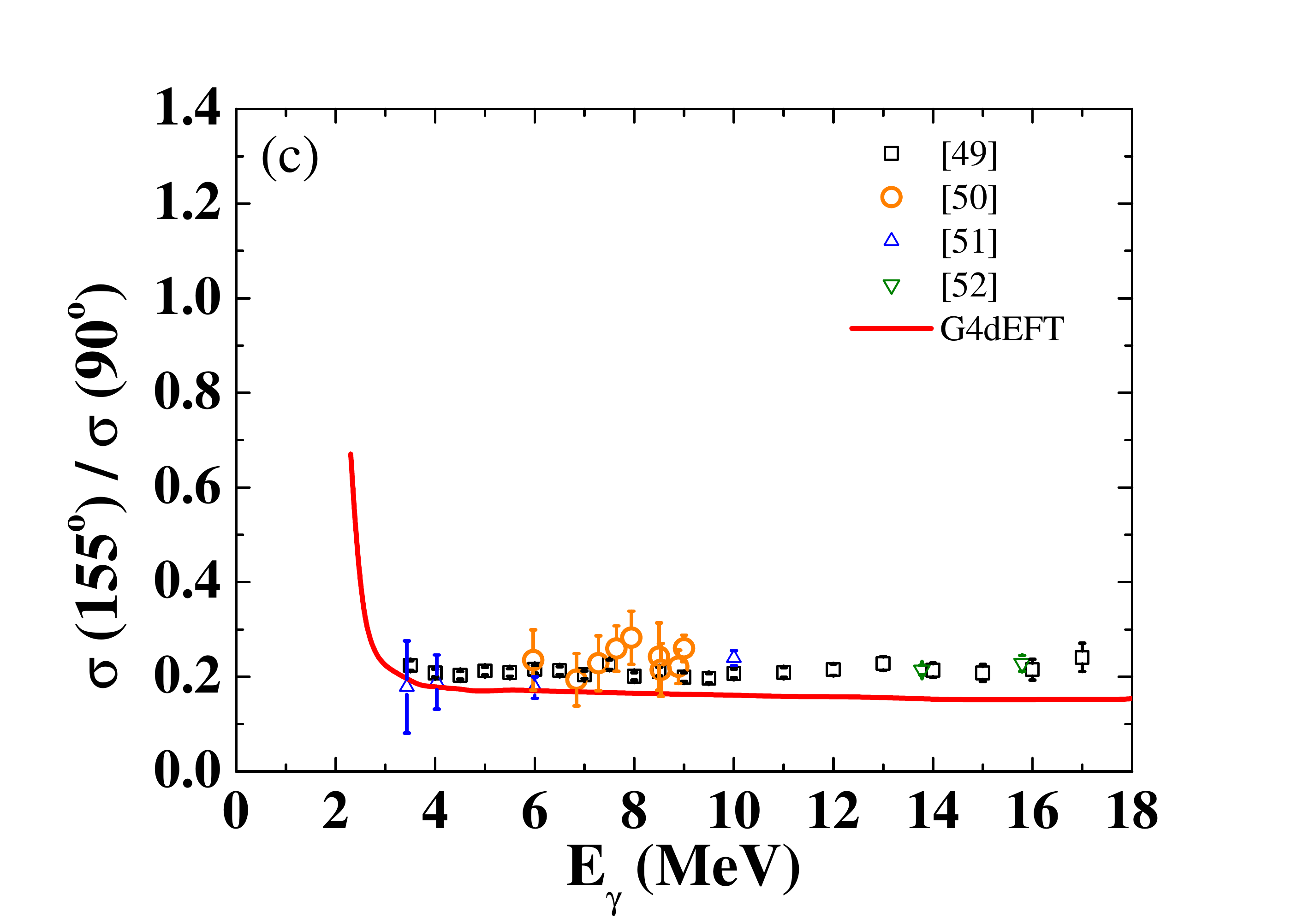, width=8cm}
\caption{(Color online) The symbols denote 
the experimental data in 
Refs. 49-52, 
and the actual numbers are taken from the EXFOR database \cite{exfor}.}
\label{figure6}
\end{figure}
%
Some experimental works report 
the ratio of differential cross sections by taking $\theta_{\rm lab} = 90^\circ$
as a reference angle \cite{rel_1, biren1988}.
These works provide data sets independent of 
the differential cross-sections 
in Figs. \ref{figure4} and \ref{figure5}.
In Fig. \ref{figure6}, 
we show differential cross-sections at $\theta_{\rm lab}=45^\circ$, $135^\circ$, and $155^\circ$
divided by those at $\theta_{\rm lab}=90^\circ$.
Experimental data are denoted with symbols, 
and G4dEFT results are depicted in solid lines.
On the average, 
our results deviate from the data of Stephenson {\it et al.}, 
Ref. 49 
by about 22\%.
Comparing the G4dEFT results to the data 
by Birenbaum {\it et al.}, 
Ref. 50, 
we have about 6\% discrepancy in the energy range 
7 $<$ $E_\gamma$ $<$ 9 MeV for $\theta_{\rm lab} = 45^\circ$,
but similar disagreement as that with Stephenson {\it et al.} 
appears at other angles and energies. 
G4dEFT cross-sections underestimate the experimental data from 
Ref. 50 
by about 15\% and 31\% for $\theta_{\rm lab} =135^\circ$ 
and $\theta_{\rm lab} =155^\circ$ on average, respectively.
On the other hand, 
a comparison with the data of Sawatzky \cite{sawa} 
gives relatively good agreement 
at energies $E_\gamma$ $\leq$ 6 MeV. 
Comparison with the data of Blackston \cite{black} shows
agreement at $\theta_{\rm lab} = 45^\circ$ with about 4.3\% error,
but the disagreements at $\theta_{\rm lab} =135^\circ$ and $155^\circ$ 
are not resolved. 

\subsection{Neutron Yields from CD$_2$ Photodisintegration}

So far, we have focused on checking the 
applicability of G4dEFT by comparing our results for $d\gamma \to np$
to experimental data.
As an application of G4dEFT, 
we attempt to predict the neutron yields from the 
$\gamma + {\rm CD}_2$ process, which has not yet been explored 
either experimentally or theoretically.
In this process, neutrons can be emitted from both carbon and deuteron.
As shown in the $d\gamma \to np$ results in Fig. \ref{figure2}, 
different results can be anticipated from G4 and G4dEFT.
One can think of gaseous (D$_2$) or liquid (D$_2$O) targets,
but taking into account the advantages of the solid target, 
such as easy handling and
large reaction rates, we chose CD$_2$ as a simulation material.

\begin{figure}[tbp] 
\centering
\epsfig{file=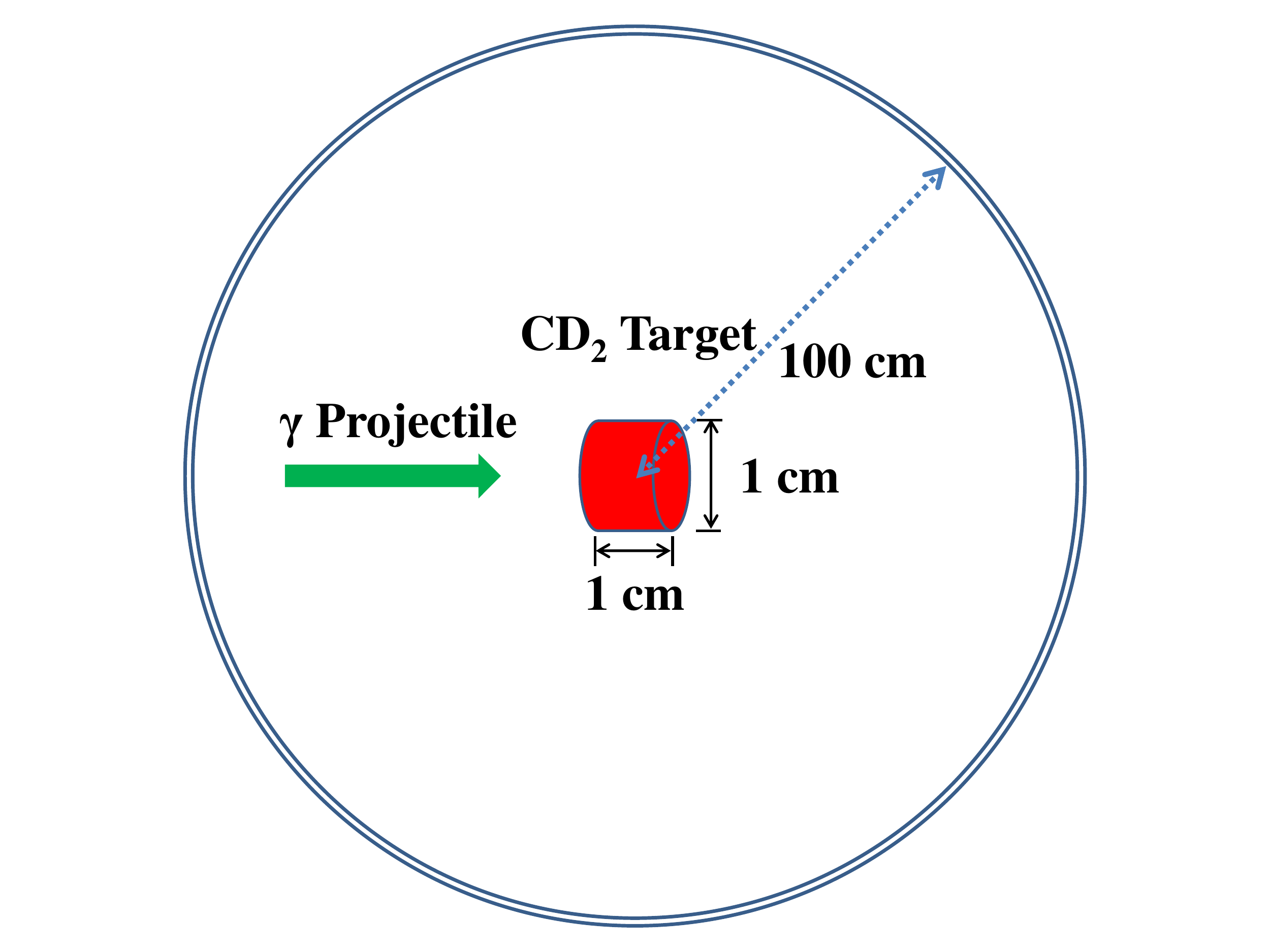, width=7cm}
\caption{(Color online) Schematic diagram of the simulation setup.}
\label{figure7}
\end{figure}

A schematic geometrical setup of our simulation 
is depicted in Fig. \ref{figure7}. 
We assume a pencil beam for the incident photons.
The CD$_2$ target 
with a density of 1 g/cm$^{3}$ 
is modeled 
as a cylinder 1 cm in diameter 
and 1 cm in thickness. 
The scoring region with the shape of a spherical shell 
surrounds the target.
The inner and the outer radii of the scoring 
region are chosen to be 100 cm and 100.1 cm, respectively.
The target area is in vacuum, and 
the thickness of the scoring region 
is chosen as 0.1 cm arbitrarily for convenience 
and does not affect the results. 

For the EM processes and hadronic interactions of hadrons 
in our simulations, 
we use GEANT4 physics lists, 
``G4EmLivermorePhysics" and ``G4HadronPhysicsQGSP$\_$BIC$\_$HP" classes, 
respectively.\footnote{\protect
There are many different pre-defined physics constructors in GEANT4. 
To check the dependence on the constructors, 
we simulated both the photon flux in the CD$_{2}$ target 
and the neutron yields generated by 
$d\gamma \to np$ reaction with different GEANT4 physics constructors, 
separately, 
but the differences due to the constructors were almost negligible 
because the simulated observables discussed in this work 
came from the $d\gamma \to np$ reaction by using G4dEFT.
} 
As mentioned in Sec. II, 
GEANT4 uses G4EmExtraPhysics for photonuclear interactions. 
G4EmExtraPhysics can be applied to carbon materials, 
but as we have shown,
it is not valid for the deuteron in the low-energy region. 
To see the effect of the deuteron, 
we simulate the $\gamma + {\rm CD}_2$ process by considering two cases. 
One is the case using only G4EmExtraPhysics (G4), 
and the other is the case using both G4EmExtraPhysics and G4dEFT (G4 + G4dEFT). 

%
%
\begin{figure}[tbp] 
\centering
\epsfig{file=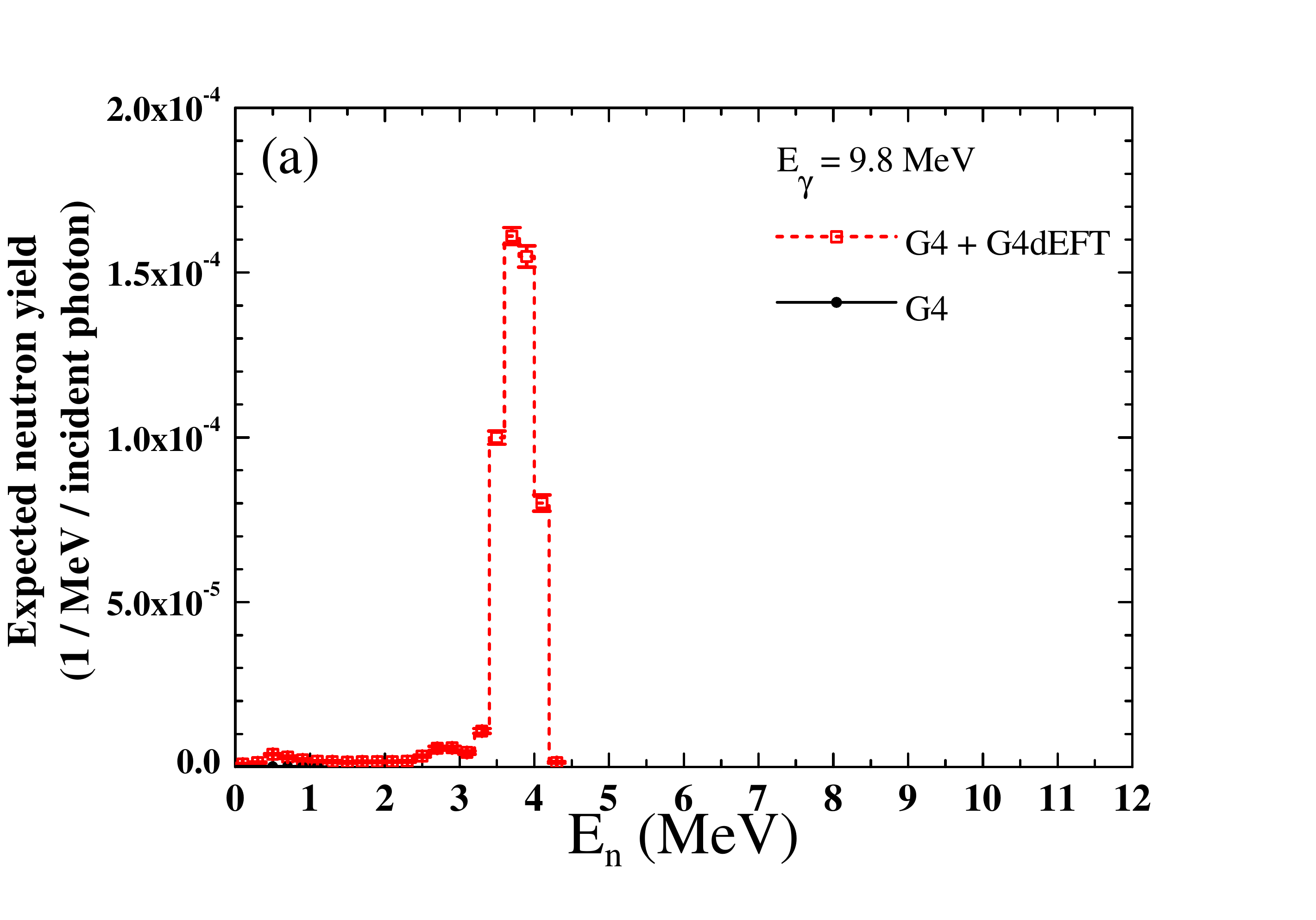, width=8cm}
\epsfig{file=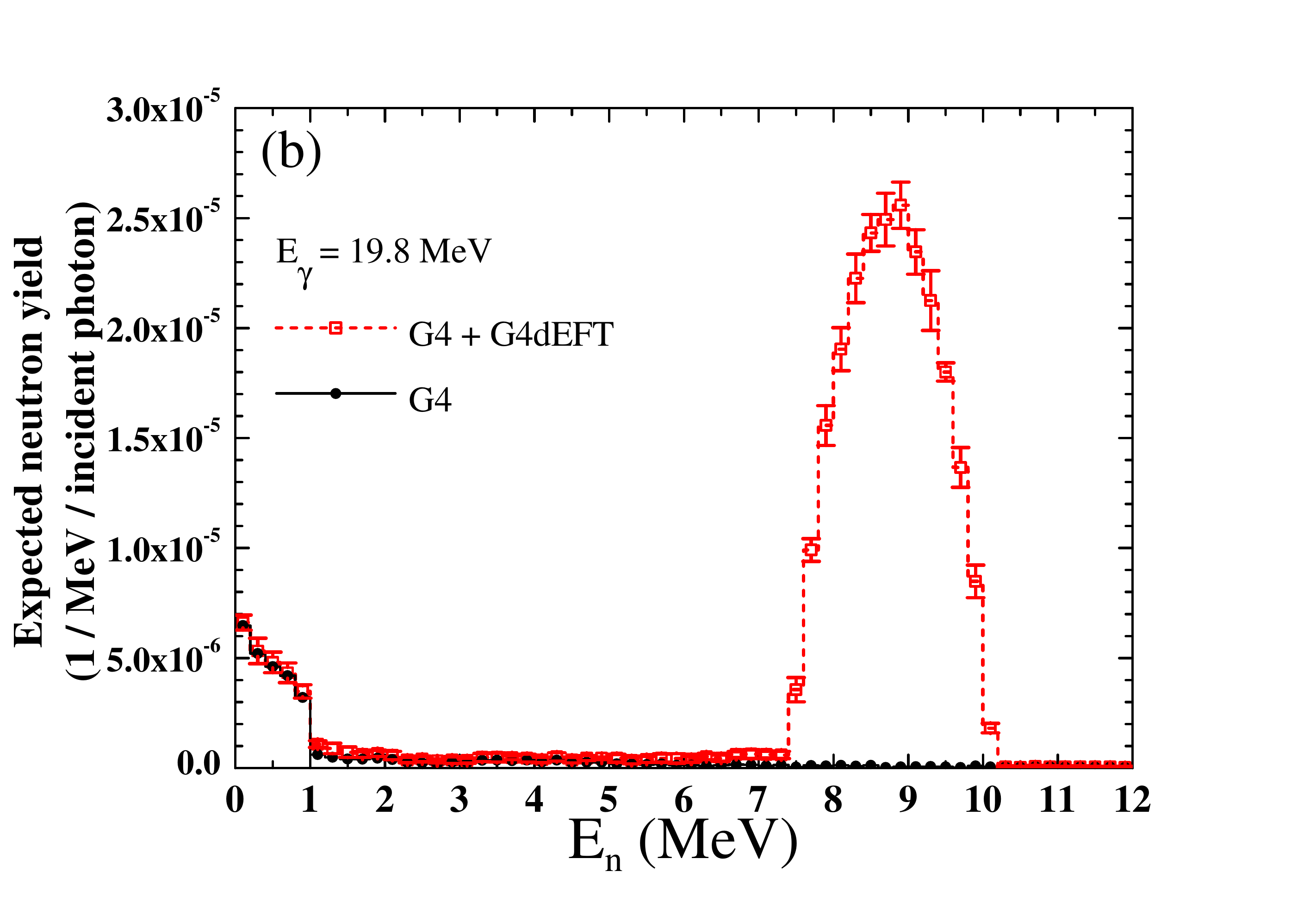, width=8cm}
\caption{(Color online) 
Expected energy spectra of neutron yields 
due to photons of energies (a) 9.8 MeV and (b) 19.8 MeV 
incident on a ${\rm CD}_2$ target. 
The black solid line with filled circles 
and the red dotted line with open squares 
represent the results obtained by using GEANT4 with G4 and G4 + G4dEFT, respectively.}
\label{figure8}
\end{figure}

Figure~\ref{figure8} shows the distribution of neutrons scored 
in the shell as a function of the neutron energy $E_n$ for (a) $E_\gamma =9.8$ MeV,
and (b) $E_\gamma = 19.8$ MeV.
In each plot, the results of G4 (G4+G4dEFT) are denoted as filled circles (open squares).
Let us analyze the results of G4 first.
As shown in Fig.~\ref{figure2}, 
the cross section of $d\gamma \to np$ is null in G4,
so the neutrons for G4 in Fig.~\ref{figure8} 
are emitted only from carbon nuclei in CD$_2$.
For $E_\gamma = 9.8$ MeV, very small numbers of neutrons 
are emitted up to $E_n \sim 1$ MeV,
but more energetic neutron are not produced. 
In the target, we assume natural carbon, $^{\rm nat}$C,
in which about 99\% is $^{12}$C, 
and the remaining 1\% is mostly $^{13}$C.
The threshold for the $^{12}$C($\gamma$,n) reaction is 18.737 MeV,
so the black circles in Fig.~\ref{figure8} (a) denote the neutrons emitted from $^{13}$C only.
Because the fraction of $^{13}$C is very small in nature, 
the production of neutrons is also very small.
Due to energy conservation, the maximum energy of the emitted neutrons is limited.
This is the reason the neutron production is truncated 
at $E_n \sim 1$ MeV.
For $E_\gamma = 19.8$ MeV, 
the photon energy exceeds the threshold 
of the $^{12}$C($\gamma$,n) reaction,
so we have a substantial number of neutrons in the energy range $E_n = 0 \sim 1$ MeV.
The number of neutrons drops abruptly around $E_n \sim 1$ MeV, 
which is also due to
the conservation of the energy of the neutrons emitted from $^{12}$C.
The black circles indicate the number 
of neutrons emitted from $^{13}$C 
at energies above $E_n \sim 1$ MeV. 
Here again, the small numbers in the range $E_n \geq 1$ MeV 
reflect the small fraction of $^{13}$C in nature.

Now, let us discuss the results of G4+G4dEFT.
In this model, the neutrons are contributed by both carbon and deuteron.
The neutron yield in the G4+G4dEFT model 
is similar to that in the G4 model in the ranges $E_n \leq 1$ MeV
for $E_\gamma = 9.8$ MeV and $E_n \leq 7.4$ MeV for $E_\gamma = 19.8$ MeV.
However, 
significant differences occur around $E_n \sim 3.7$ MeV 
for $E_\gamma =9.8$ MeV 
and around $E_n \sim 8.7$ MeV for $E_\gamma = 19.8$ MeV,
where high and broad peaks appear.
The number of neutrons produced by the $d\gamma \to np$ reaction 
is obtained by making use of the ``G4UserSteppingAction" class. 
The ratio of the neutron number for $E_\gamma =9.8$ MeV to that for $E_\gamma = 19.8$ MeV
is obtained as $2.293 \pm 0.046$. 
This number is close to the ratio of the total cross-sections 
of the $d\gamma \to np$ process, 
2.287, for $E_{\gamma}$ = 9.8 MeV (1.44 mb) 
and for $E_{\gamma}$ = 19.8 MeV (0.63 mb). 
The similarity of two ratios is proof that 
the peaks in Fig.~\ref{figure8} originate from interactions of photons and deuterons.

\begin{figure}[tbp] 
\centering
\epsfig{file=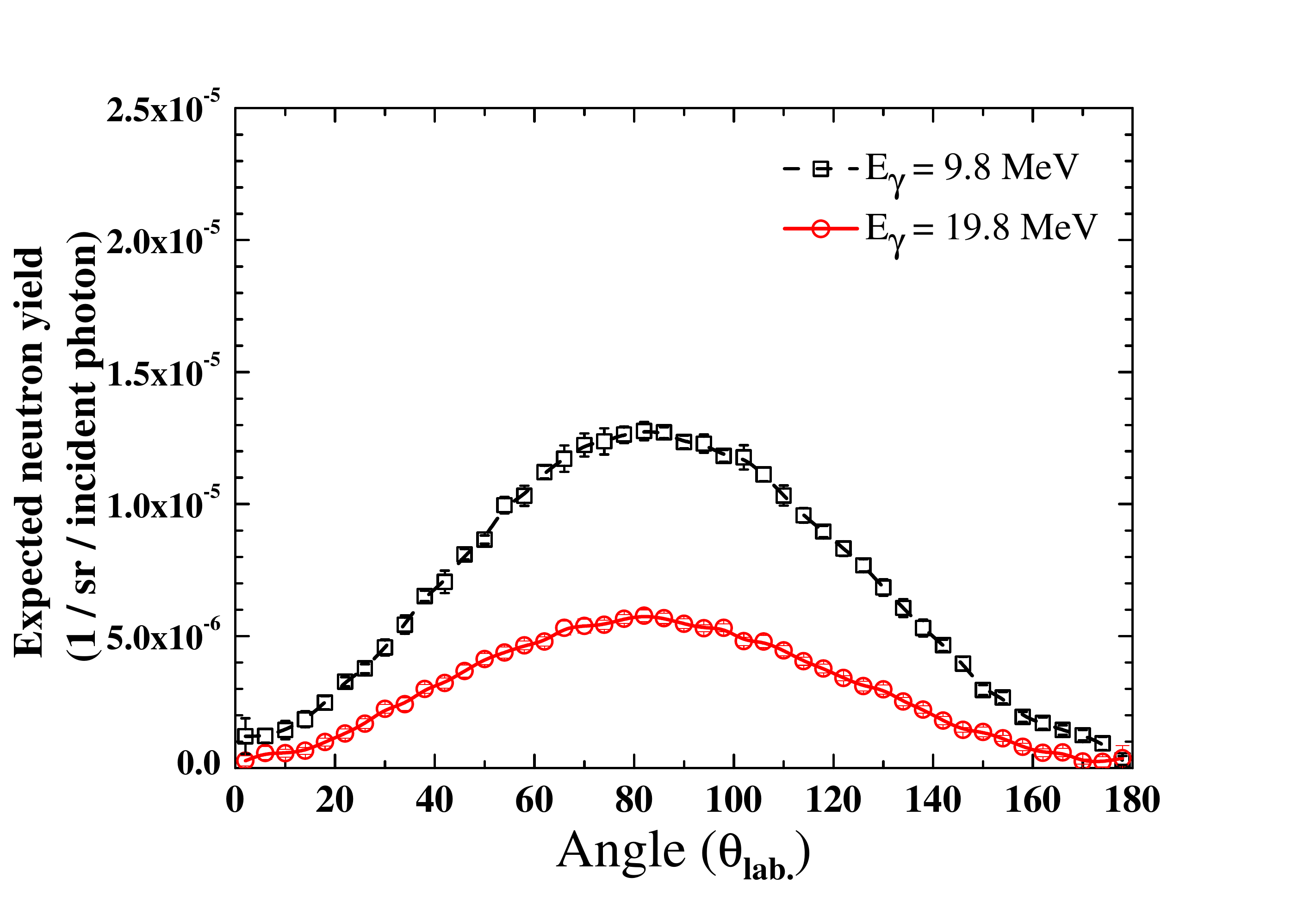, width=8cm}
\caption{(Color online) 
Expected angular distributions of neutron yields in the G4+G4dEFT model. 
The black dotted line with squares and 
the red solid line with circles represent the results 
for 9.8 MeV and 19.8 MeV photons, respectively.}
\label{figure9}
\end{figure}

In Fig. \ref{figure9}, 
we show the angular distribution of the neutron yield.
The maximum values are located around $\theta_{\rm{lab}}$ = $90^\circ$,
which is in agreement with the behavior of 
the differential cross-section of 
the $d\gamma \to np$ process, Fig.~\ref{figure4}.
This coincidence proves again that most of the neutrons in the peaks are emitted
from the deuterons in CD$_2$.
The angular distribution for $E_\gamma = 19.8$ MeV is relatively broader 
than that for $E_\gamma=9.8$ MeV, which is one of the reasons 
we have a peak broader for $E_\gamma =19.8$ MeV than 
for $E_\gamma = 9.8$ MeV
in Fig.~\ref{figure8}.

\begin{figure}[tbp] 
\centering
\epsfig{file=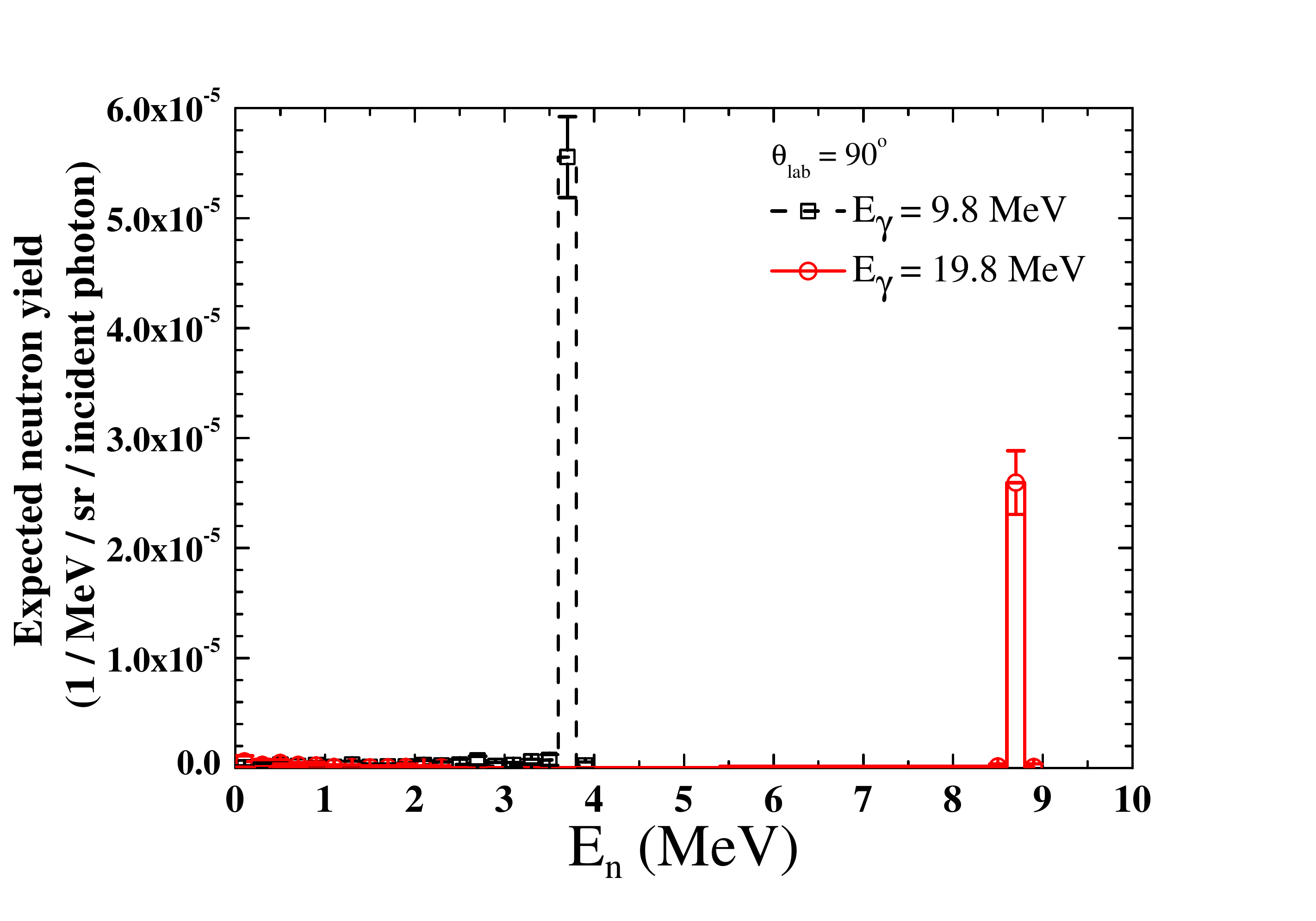, width=8cm}
\caption{(Color online) 
Expected neutron energy spectra 
at ${\rm{\theta}}_{\rm{lab}}$ = $90^\circ$ 
produced by photons on a ${\rm CD}_2$ target
in the G4+G4dEFT model.}
\label{figure10}
\end{figure}

The distribution of the emitted neutrons in terms of the energy at 
$\theta_{\rm{lab}} = 90^\circ$ is plotted in Fig.~\ref{figure10}. 
Very sharp peaks are seen at $E_n \sim 3.7$ MeV for $E_\gamma = 9.8$ MeV 
and at $E_n \sim 8.7$ MeV for $E_\gamma = 19.8$ MeV.
If one treats the scattering process $d\gamma \to np$ classically and 
calculates the energy of outgoing neutrons in the laboratory frame,
one can easily obtain $E_n = 3.7$ MeV and 8.7 MeV for $E_\gamma = 9.8$ MeV and
19.8 MeV, respectively.
The coincidence of these values and the positions of the peaks 
signal the dominance of $d\gamma$ reactions 
in the emission of neutrons.

\section{Summary}

With the development of effective field theories in nuclear physics,
the accuracy in describing and understanding few-nucleon systems 
is being improved unprecedentedly.
The improvement will make it possible to have more precise results 
for simulations in which few-nucleon processes will contribute non-negligibly.
Our work started from the question of how we could embed the progress 
of nuclear theory for few-nucleon systems in the simulations of nuclear processes.
As a first step, 
we checked the validity of 
G4EmExtraPhysics, which accounts
for the photonuclear interactions in GEANT4 (v10.1) 
by calculating the total cross-section
of the $d\gamma \to np$ process.
We found that the G4EmExtraPhysics model gave null result for
the total cross-section of the process 
at energies below the pion threshold.
In order to resolve the problem and enhance the predictive power of simulations, 
we incorporated analytic results of a dibaryon effective field theory into GEANT4.

The new model has been checked thoroughly by calculating the total and
the differential cross-sections of the $d\gamma \to np$ process 
for incident photon energies up to 70 MeV 
and by comparing the results with available experimental data.
G4dEFT gives results within the experimental error bars 
for $E_{\gamma}$ below 15 MeV. 
For photon energies above 15 MeV, 
the results from G4dEFT show reasonable agreement 
with those from experiments within 20\%, 
even though the energy region is 
out of the valid range for the theory. 

The model has been applied to calculate the yield 
of neutrons emitted from CD$_2$ bombarded by photons.
Because no experimental datum is available for the observable,
only a comparison between the new and the existing
models was made.
The two models show a critical difference 
in the prediction of the neutron yield.
The measurement, if performed, will provide criteria about 
the accuracy and the extent to which the model is valid and applicable.

\begin{acknowledgments}
This research was supported by the Daegu University Research Grant, 2013.
\end{acknowledgments}

\end{document}